\documentclass[useAMS]{mn2e}
\usepackage{graphicx}
\usepackage{latexsym}
\usepackage{epsfig}

\title{Synthetic observations of simulated pillars of creation}

\author[Barbara Ercolano, Jame E. Dale, Matthias
Gritschneder, Mark Westmoquette]{Barbara Ercolano$^{1,2}$\thanks{E-mail: ercolano@usm.lmu.de (BE)}, James
  E. Dale$^{2}$, Matthias Gritschneder$^{3}$, Mark Westmoquette$^{4}$\\
$^1$Universit\"ats-Sternwarte M\"unchen, Scheinerstr. 1, 81679 M\"unchen, Germany\\
$^2$Cluster of Excellence ’Origin and Structure of the Universe’, Boltzmannstr.2, 85748 Garching, Germany\\
$^3$Kavli Institute for Astronomy and Astrophysics, Peking University,Yi He Yuan Lu 5, Hai Dian, 100871 Beijing, China\\
$^{4}$European Southern Observatory, Karl-Schwarzschild-Str. 2, 85748 Garching bei M\"unchen, Germany}

\begin{document}

\maketitle

\label{firstpage}

\def\mnras{MNRAS}
\def\apj{ApJ}
\def\aap{A\&A}
\def\apjl{ApJL}
\def\apjs{ApJS}
\def\bain{BAIN}
\def\araa{ARA\&A}
\def\pasp{PASP}
\def\aj{AJ}
\def\pasj{PASJ}
\def\ga{\sim}
\voffset=-0.5in

\begin{abstract}

We present synthetic observations of star-forming interstellar medium
structures obtained by hydrodynamic calculations of a turbulent
box under the influence of an ionising radiation field. The
morphological appearance of the pillar-like structures in optical
emission lines is found to be very similar to observations of nearby
star forming regions. We calculate line profiles as a function of
position along the pillars for collisionally excited
[OIII]$\lambda$5007, [NII]$\lambda$6584 and
[SII]$\lambda$6717, which show typical FWHM of 2--4 km s$^{-1}$
km/s. Spatially resolved emission line diagnostic diagrams are also
presented which show values in general agreement with observations of
similar regions. The diagrams, however, also highlight significant
spatial variations in the line ratios, including values that would be classically
interpreted as shocked regions based on one-dimensional
photoionisation calculations. These values tend to be instead the
result of lines of sight intersecting which intersect for large portions of their lenghts the ionised--to--neutral transition regions in the gas. We caution therefore against
a straightforward application of classical diagnostic diagrams and one--dimensional photoionisation calculations to spatially resolved
observations of complex three-dimensional star forming regions. 

\end{abstract}


\section{Introduction}
The influence of stellar feedback on the star formation process itself
is one of the most important issues in this field of
astrophysics. Feedback from OB--type stars, in the form of
photoionizing radiation, radiation pressure, winds, and supernovae,
can be both positive (i.e. enhancing or hindering the star formation
efficiency) and negative and may be responsible for regulating the efficiency and rate of star formation on the scale of giant molecular clouds (GMCs).\\
\indent The destructive effects of feedback, in the sense of quenching star formation, gas--expulsion and dispersal of young clusters, have been studied by many authors. The dynamical influence of gas expulsion on clusters has been examined by, e.g., Hills (1980), Goodwin (1997), Boilly \& Kroupa (2003,a,b) Goodwin \& Bastian (2006), while self--regulation of star formation by limiting the star formation efficiency has been modelled by, e.g., Whitworth (1979), Bodenheimer et al. (1979), Tenorio--Tagle \& Bodenheimer (1988), Franco et al. (1994), Matzner et al. (2002).\\
\indent Conversely, numerous authors have studied the positive effects
of feedback in the context of induced or triggered star formation,
partly spurred on by a recent crop of exquisite observations of
triggering apparently in progress. The popular collect--and--collapse
model of triggering has been studied analytically and numerically many
times (e.g. Whitworth et al., 1994, W\"unsch \& Palou\v{s}, 2001, Dale
et al., 2007a) and has been observed in action (e.g. Zavagno et al,
2006, Deharveng et al., 2008, Zavagno et al., 2010). However, for this
model to be valid, it is necessary to assume that the gas on which feedback is acting is relatively smooth and quiescent, conditions which are not always met, especially not in the inner regions of turbulent GMCs where the density and velocity fields in the gas are highly inhomogeneous. Under these conditions, it is very difficult to devise an analytic description of the effects of feedback.\\
\indent This question has instead been approached numerically by means of
hydrodynamical simulations modified to include the thermal effect of
photoionisation. The complexity of such simulations and the inevitable
dependence of the results on the chosen initial conditions demand
strict comparisons with observations before meaningful conclusions
may be drawn. 
One such test for how realistic these models are involves running large-scale cluster
simulations and computing statistics for (e.g.) the initial mass
function (IMF) of the stars and their multiplicity. Such statistics,
however, are often difficult to obtain with the required accuracy. The
stellar IMF is a steep function and at least several hundred stars are
required to adequately sample it over a meaningful range of
masses. Combined with the difficulty of modelling feedback itself,
this problem requires large and time--consuming calculations (
  Klessen, Krumholz \& Heitsch 2009; Dale and Bonnell, 2011, submitted).\\ 
\indent Another approach is to compare the morphology of structures in the
interstellar medium (ISM) that are influenced by the effects of
photoionising radiation. Spatially resolved imaging (e.g. from the Hubble Space Telescope) and spectroscopy of young star forming regions (e.g.\ Hester et al. 1996; Pound et al. 2003; Gahm et al. 2006; Matsuura et al. 2007; Smith et al. 2010; Westmoquette et al. 2010), provide a wealth of crucial information and hold the promise of helping to constrain models of ISM evolution.
Until now, the problem of realistically computing the observable radiation fields 
from systems produced by hydrodynamical calculations has only been moderately
explored (Arthur et al 2001). The main reason for this is that
the majority of hydrodynamical codes only include the thermal effects
of photoionisation in  a very approximate fashion (see Ercolano \&
Gritschneder, 2011a; Dale et al 2007b), preventing the calculation of the detailed
thermal and ionisation structure of the gas  required for the
prediction of emission line and continuum radiation from the region. A
solution to this problem is to employ a dedicated three-dimensional
photoionisation code as a postprocessing step to analyse the emission
from density snapshots of the hydrodynamical calculations. 

We use this approach in this paper to explore the observational
appearance of the star-forming pillars obtained by the recent
calculations of Ercolano \& Gritschneder (2011b, EG11), which modelled the impact of radiation from high--mass stars on the turbulent ISM. The calculations
improved upon the previous work of Gritschneder et al. (2009b, 2010) by
including the effects of diffuse ionizing radiation fields on the hydrodynamics  and thus on the
shaping of the ISM and the formation of pillar-like structure. EG11 found that the diffuse
fields promote the detachment of dense clumps and filaments from the
parent cloud, and produce denser and more spatially confined
structures. In this paper we wish to investigate the observational
appearance of such structures in optical emission, to verify the
morphological similarity to observations, and to explore the potential of
spatially resolved line ratio diagnostics.

We describe our numerical methods in Section 2 and main results in Section
3. Section 4 contains a brief summary and conclusions. 

\section{Numerical Methods}

\subsection{Hydrodynamical calculations}
\begin{figure*}
\begin{center}
\includegraphics[width=\textwidth]{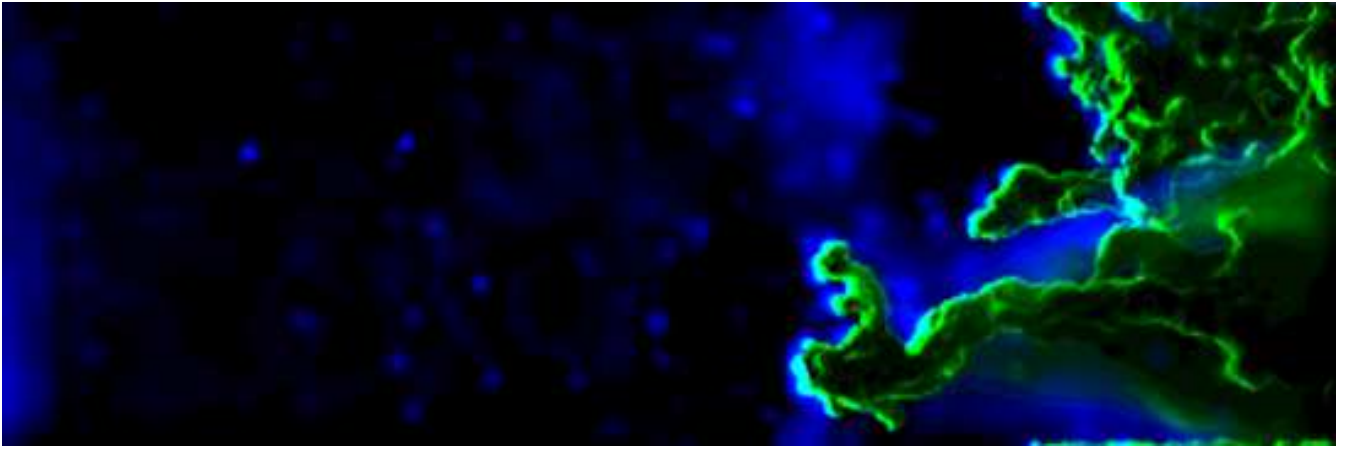}
\caption{False colour
composite image of the EG11 pillar at t = 500kyr, where red is
H$\alpha$, blue is [OIII]$\lambda\lambda$5007,4959 and green is a
combination of the two lines.}
\end{center}
\end{figure*}

The hydrodynamic simulations were performed with the tree-SPH/ionisation
code iVINE (Gritschneder et al 2009a). iVINE treats the ionisation of
the turbulent ISM under the assumption of plane-parallel irradiation
on the computational domain. The surface facing the ionising source is
decomposed into equally--spaced bins, whose size is chosen
according to the smoothing length of the particles there. The bins or rays
are subsequently refined as the radiation penetrates the gas according to the local smoothing
length. Along these rays the optical depth is calculated and a new
pressure is assigned to each particle $i$ according to: 
\begin{equation}
P_\mathrm{i} = \left(\frac{T_\mathrm{hot} \eta_\mathrm{i}}{\mu_\mathrm{hot}} +
  \frac{T_\mathrm{cold} (1-\eta_\mathrm{i})}{\mu_\mathrm{cold}}\right)
\frac{k_\mathrm{B} \rho_\mathrm{i}}{m_\mathrm{p}}.
\end{equation}
Here, $\eta$ is the ionisation degree, defined as the ratio of the
number density of free electrons to the total number of hydrogen nuclei, 
 $T_{hot}=10^4K$, $T_{cold}=10K$, 
$\mu_\mathrm{hot}=0.5$ and $\mu_\mathrm{cold}=1.0$ are the temperatures
and the mean molecular weights of the ionised and 
neutral gas in the case of pure hydrogen,
respectively. $k_\mathrm{B}$ is the Boltzmann constant, $m_\mathrm{p}$
is the proton mass and $\rho_\mathrm{i}$ is the density of the SPH 
particle. 

In the results presented here, the diffuse ionisation is included
according to EG11. The shadowed particles with a density lower than
$n=100cm^{-3}$ are assigned a new, higher temperature according to a
fitted function. This function is derived from comparisons of previous
simulations with full 3D postprocessing using the {\sc mocassin} code
(Ercolano et al. 2003, 2005, 2008). The 
cold, dense, neutral gas is kept at a temperature of $T_{cold}=10K$
under the assumption that it can cool fast enough, i.e. it is treated as 
isothermal. 

Hydrodynamic and gravitational forces are then calculated for the
particles with the accuracy parameters given in (Gritschneder
2009b). The initial density and velocity distribution mimics the
turbulent ISM at Mach 5 inside a computational domain of $4pc^3$ with
a mean density of $n=300cm^{-3}$. To create synthetic observations of
the resulting structures discussed in EG11, we take a subdomain
starting from the irradiated surface at the $y$=0 plane and spanning $3pc$ in the $y$-direction
and $1pc$ in the $x$- and $z$-directions around the most prominent
features of the simulations in EG11. We map the hydrodynamic
quantities in this region on an equally--spaced grid
($384\times128^2$). With this data we perform the postprocessing with
{\sc mocassin} described below. 

\subsection{Photoionisation calculations}

We used the 3D photoionisation code {\sc mocassin} (Ercolano et
al. 2003, 2005, 2008) to compute the
detailed ionisation and temperature structure from the 5kyr snapshot
of the above--described region containing the main pillar in the EG11 simulation. The code uses a Monte Carlo
approach to the transfer of radiation, allowing the treatment of both the
direct stellar radiation and the diffuse fields for arbitrary
geometries and density distributions. The gas is heated mainly by
photoionisation of hydrogen and other abundant elements. Cooling
is dominated by collisionally excited lines of heavy elements, but
contributions from recombination lines, free-bound, free-free and
two-photon continuum emission are
also included. We have
assumed typical HII region abundances for the gas phase, namely (given as 
number densities with respect to hydrogen): He/H = 0.1, C/H = 2.2e-4, N/H
= 4.0e-5, O/H = 3.3e-4, Ne/H = 5.0e-5, S/H = 9.0e-6. 

Emission lines produced in each volume element are then computed
by solving the statistical equilibrium problem for each atom and ion at
the local temperature and ionisation conditions. The atomic database
includes opacity data from Verner et al. (1993) and Verner \& Yakovlev
(1995), energy levels, collision strengths and transition
probabilities from Version 5.2 of the CHIANTI database (Landi et
al. 2006, and references therein) and the hydrogen and
helium free-bound continuous emission data of Ercolano \& Storey
(2006). 

The current version of the code can treat the atomic and ionised
gas phases and dust, but no chemical network is currently included,
preventing the calculation of physical conditions and emission from
photodissociation or molecular regions (but see Bisbas et al 2011 for
recent developments in this area). 

\subsection{Visualisation Tool}

The 3D grids of temperature, ionisation structure and emission from
{\sc mocassin} were further post-processed by means of our newly--developed, 
Python-based suite of tools, which will shortly be made freely available to the community
on the {\sc mocassin} website (www.3d-mocassin.net). 
The tools provide 2D projections of the grids along the three major
axes, accounting for dust extinction along the line of sight. Velocity maps,
line profiles and emission line diagnostic diagrams, like those
presented in Section 3 of this paper, are also
available. The new software will be distributed with an online manual,
where further details on the available routines will be provided. 

\begin{figure*}
\begin{center}
\includegraphics[width=18cm]{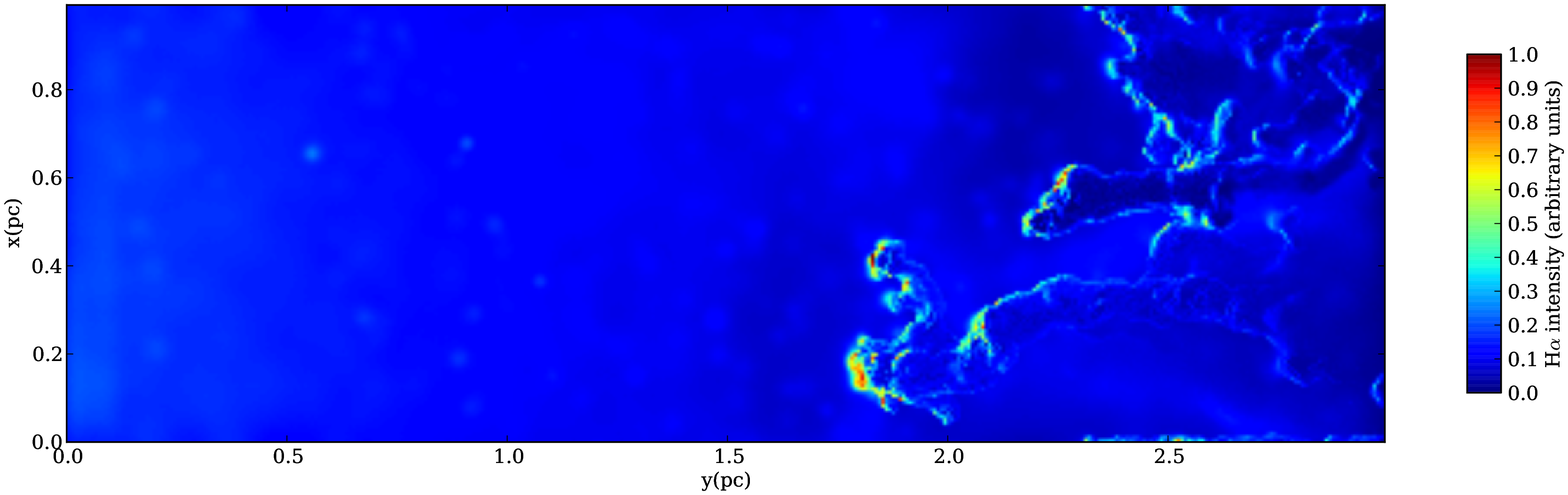}
\includegraphics[width=18cm]{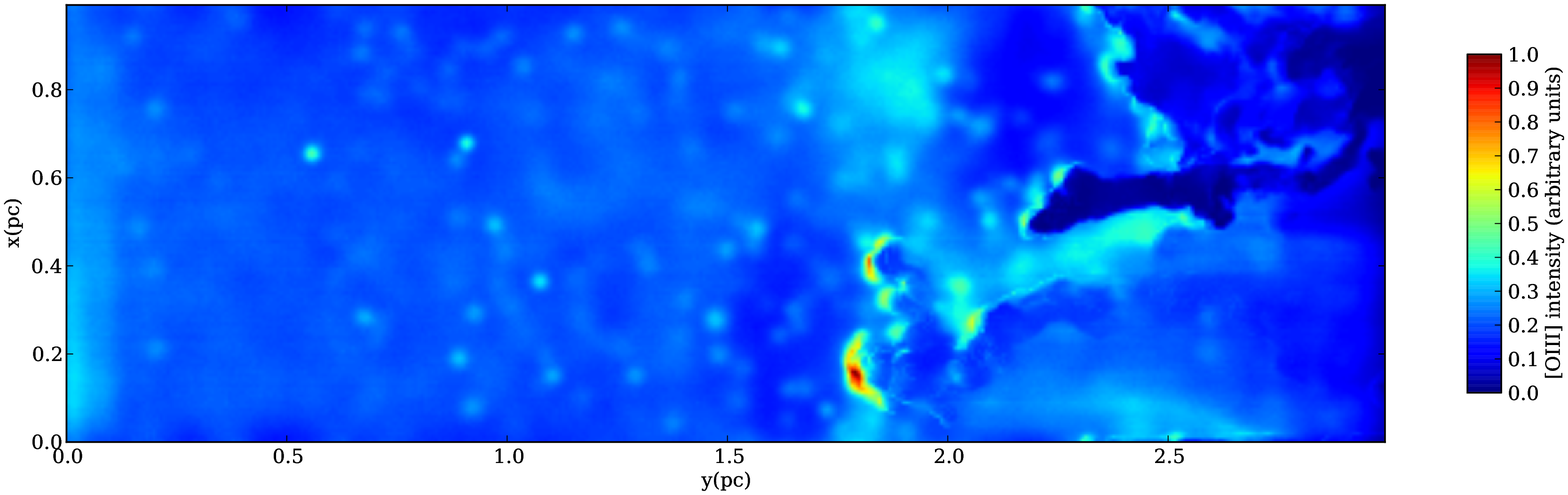}
\caption{Top Panel:  H$\alpha$ image of the EG11 pillar at t =
  500kyr. Bottom Panel: same image in [OIII]$\lambda\lambda$5007,4959.}
\end{center}
\end{figure*}

\section{Results}

\begin{figure*}
\begin{center}
\includegraphics[width=0.4\textwidth]{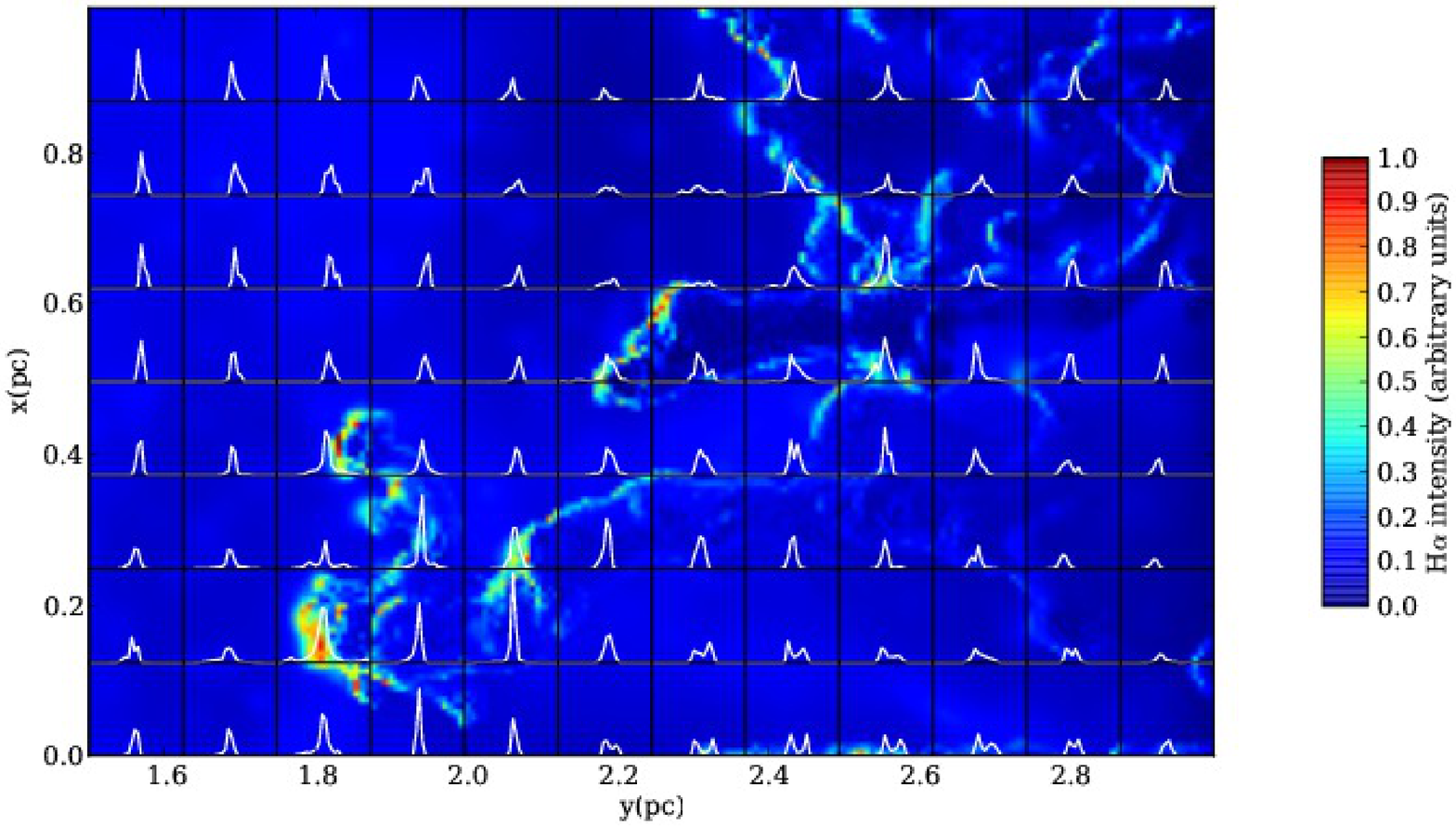}
\includegraphics[width=0.4\textwidth]{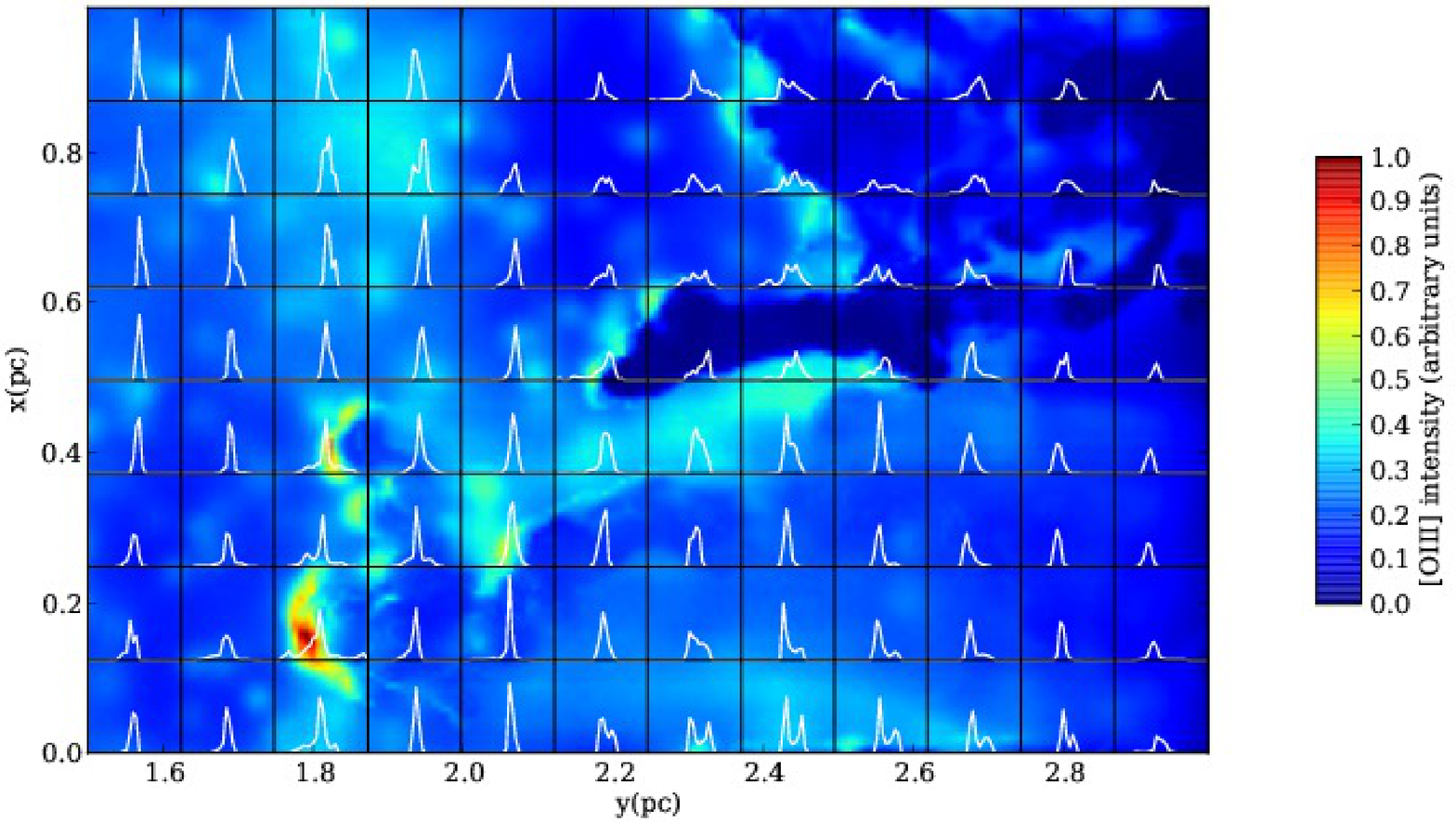}
\includegraphics[width=0.4\textwidth]{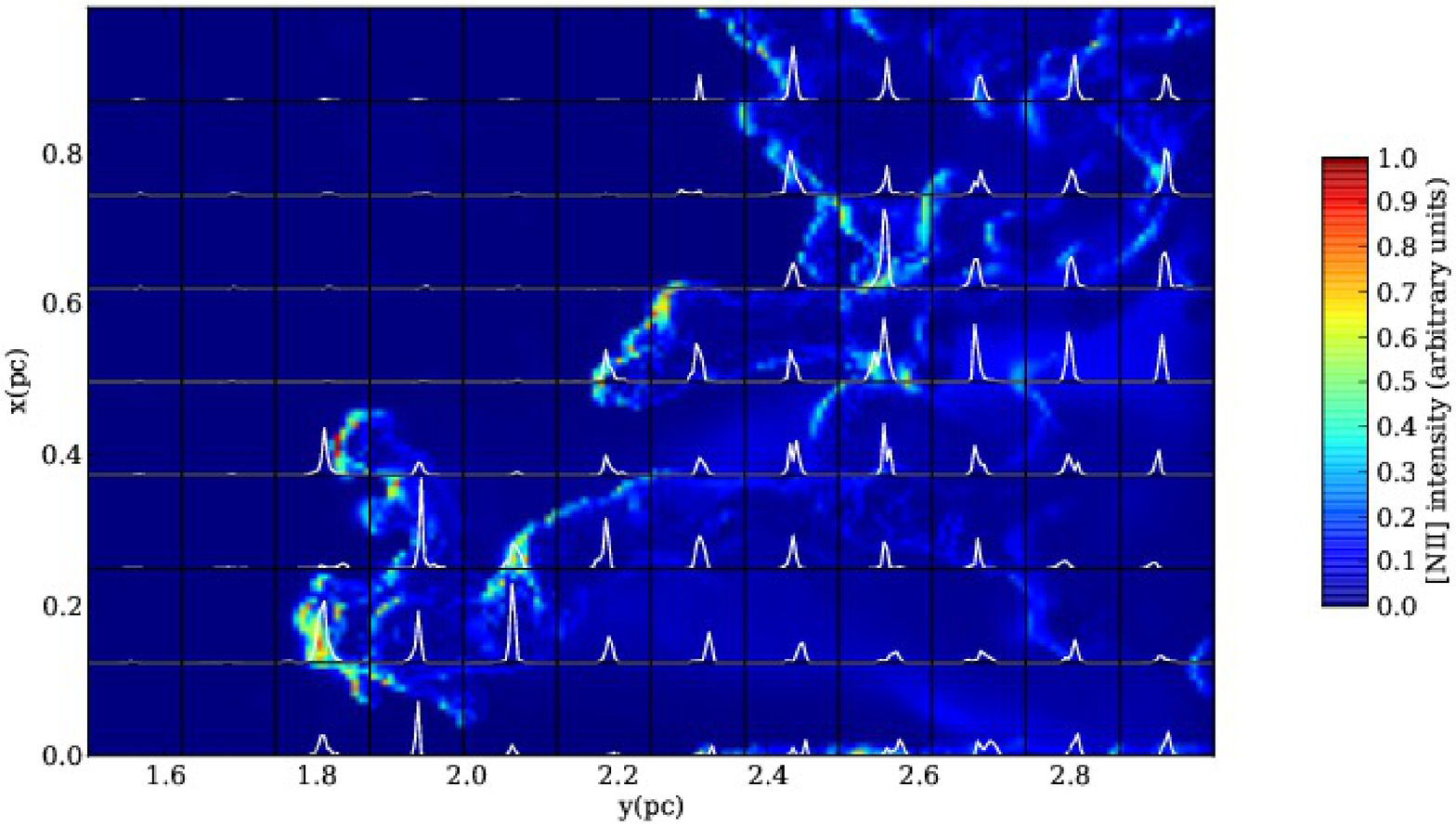}
\includegraphics[width=0.4\textwidth]{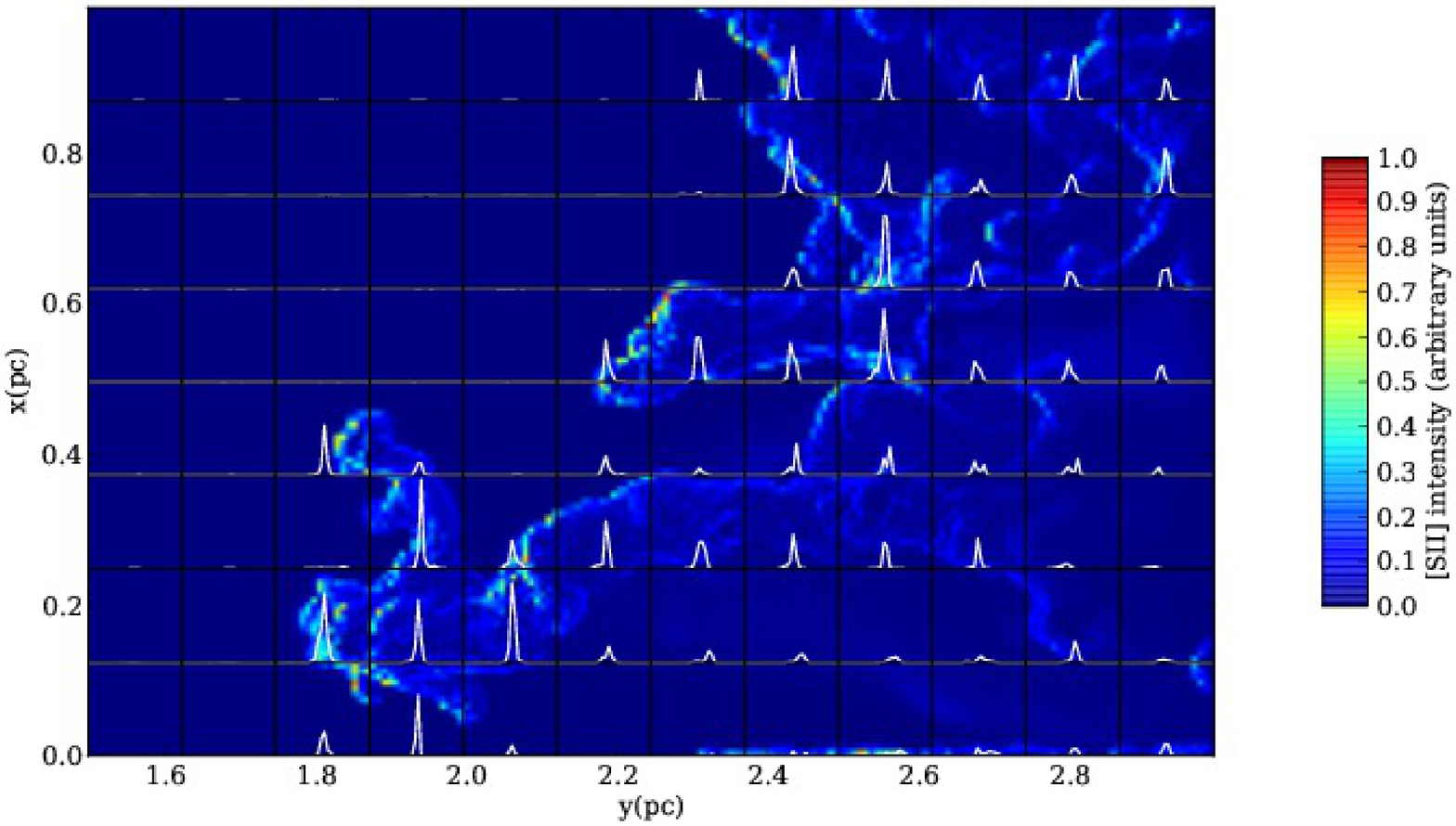}
\caption{Predicted
profiles for H$\alpha$ (top right), [OIII]$\lambda$5007 (top left),
[NII]$\lambda$6584 (bottom left) 
and [SII]$\lambda$6717 (bottom right) are shown. The width of each box
is 30 km s$^{-1}$ (centred on zero) and the 
emissivities are in arbitrary units, normalised to the brightest line.}
\end{center}
\end{figure*}

\begin{figure}
\begin{center}
\includegraphics[width=9cm]{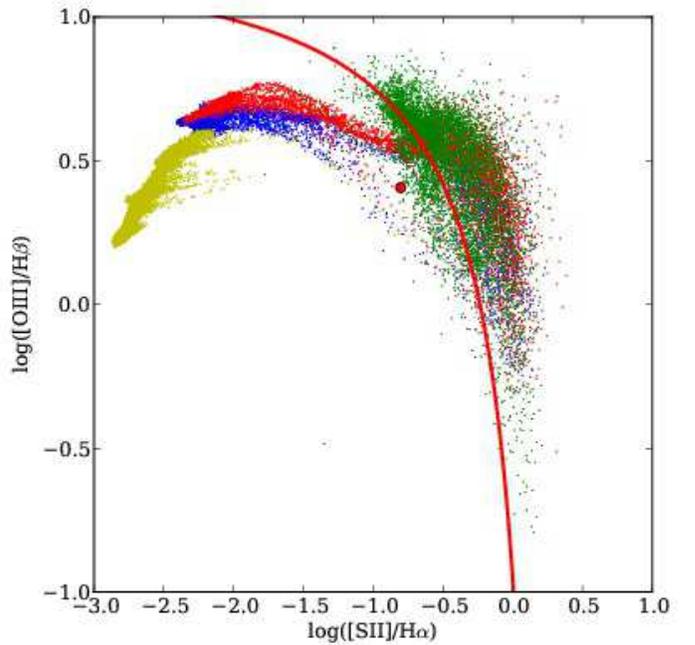}
\caption{[OIII]/H$\beta$ versus [SII]/H$\alpha$ at each pixel on the
  2D projected images. The
integrated value for all pixels is shown as the red point on the
figures. The solid red line represents the Kewley et al
(2001) line separating star--forming galaxies from AGN. The points are
colour-coded according to distance from the left-hand side of the
simulation box. The green points are for $y$ $>$ 2.6pc, red for 2.6pc $>$ $y$
$>$ 2.2pc, blue for 2.2pc $>$ $y$ $>$ 1.8pc and yellow for $y$ $<$ 1.8pc. }
\end{center}
\end{figure}

\begin{figure}
\begin{center}
\includegraphics[width=9cm]{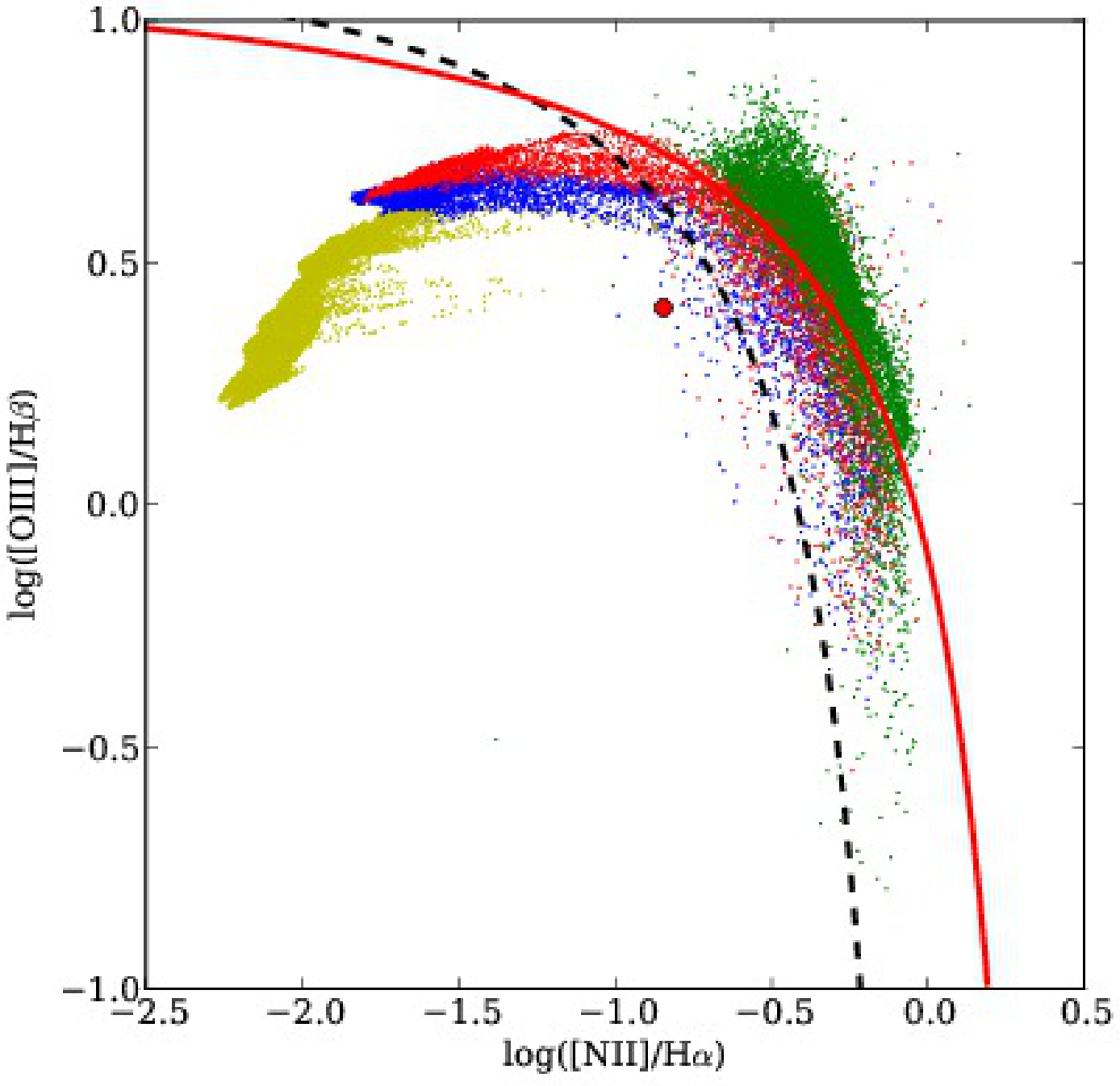}
\caption{[OIII]/H$\beta$ versus [NII]/H$\alpha$ at each pixel on the
  2D projected images. The
integrated value for all pixels is shown as the red point on the
figures. The solid red line represents the Kewley et al
(2006) line separating star--forming galaxies from AGN, while the dashed line shows the line from Kauffmann et al 2003 separating pure--star--forming galaxies from composite star--forming/AGN systems. The points are
colour-coded according to distance from the left-hand side of the
simulation box. The green points are for $y$ $>$ 2.6pc, red for 2.6pc $>$ $y$
$>$ 2.2pc, blue for 2.2pc $>$ $y$ $>$ 1.8pc and yellow for $y$ $<$ 1.8pc. }
\end{center}
\end{figure}

\subsection{Morphology}

EG11 concluded that one of the effects of diffuse fields in the simulation
of pillar formation was that the structures obtained were denser,
spatially thinner and that they would eventually detach from the
parent turbulent cloud. The famous HST images of e.g. the Pillars of
Creation in the Eagle nebula show, however, structures that are apparently coherent. We
constructed 2D projected intensity maps of optical emission lines from
the 500kyr snapshot
of the EG11 simulations to verify that the apparent morphological
constraint provided by the HST image of coherent pillars is still
respected. Figure 1 shows a combined false colour
composite image of the EG11 pillar at t = 500kyr, where red is
H$\alpha$, blue is [OIII]$\lambda\lambda$5007,4959 and green is a
combination of the two channels. The top and bottom panels of Figure~2 show the same region in the
individual channels, respectively H$\alpha$ and
[OIII]$\lambda\lambda$5007,4959. All images account for dust extinction
along the line of sight, for the length of the simulation box, according to the interstellar extinction curve
of Weingartner \& Draine (2001) for a Milky Way grain size distribution for
RV =3.1, with C/H = bC = 60 ppm in log-normal size distributions, but
renormalised by a factor 0.93. This grain model is considered to be
appropriate for the typical diffuse HI cloud in the Milky Way. 

The figure clearly shows that the line of sight integrals of
emission lines from the ionised gas at the surface of the pillars,
give the desired appearance of internally coherent structures. The
spatial thickness of such structure is of the order of a tenth of a
parsec, which is consistent
with observations.

\subsection{Line profiles}

The gas velocity information combined with the emission measure at each
volume element allows us to compute predicted emission line profiles
along the structures. This is shown in Figure~3 where the predicted
profiles for H$\alpha$ (top right), [OIII]$\lambda$5007 (top left),
[NII]$\lambda$6584 (bottom left) 
and [SII]$\lambda$6717 (bottom right) are shown. The width of each box
is 30 km s$^{-1}$ (centred on zero) and the 
emissivities are in arbitrary units, normalised
across the regions to the brightest lines. Typical FWHM for the profiles are 2--4 km s$^{-1}$. 
It is useful to note at this point that
Westmoquette et al (2009) report the presence of both a narrow
($\sim$20km/s) and a broad ($50-150$km/s) component in the 
H$\alpha$ line profiles from their optical/near-IR IFU observations of
a gas pillar in the Galactic HII region NGC 6357, which contains the
young open star cluster Pismis 24. They interpret the broad component
as being formed in ionized gas within turbulent mixing layers on the
pillar's surface, generated by the shear flows of the winds from the O
stars in the cluster. Our simulations do not include
stellar winds and are instead influenced only by the thermal pressure of the hot ionised gas. Relative gas velocities therefore cannot exceed a few times the speed of sound in the ionised gas, approximately 10km s$^{-1}$, which explains why we only can
reproduce the narrow component of the emission lines.

\subsection{Line Ratio Diagnostics}

Emission line ratios are often used as diagnostics of gas properties
such as temperature and density, and sometimes also to distinguish
between shock-- and photo--ionisation in the gas. The development of
diagnostic tools dates back to more than forty years ago when
one-dimensional photoionisation models started being
developed (e.g. Harrington 1968). Diagnostic diagrams were
successively developed, based on grids of 1D photoionisation
models, often calibrated with empirical data from spatially unresolved
observations. Some of the most widely used
diagnostic diagrams are the so-called BPT diagrams (Baldwin, Phillips
\& Terlevich 1981), which
consist of comparing the ratios of strong collisionally excited lines
(e.g. [OIII]$\lambda\lambda$5007,4959, [NII]$\lambda$6583, 48, [SII]$\lambda$6717,31) to the main
hydrogen recombination lines (e.g. H$\beta$, H$\alpha$). The excitation and ionisation parameter in a given region can be determined by 
comparison of different ionised species or of lines with different
temperature dependences. Calibration of models
with observations has allowed regions in such plots
where the line ratios are dominated by photoionisation over shocks to be distinguished, and
the determination of the nature of astronomical objects, e.g. HII regions, active
galactic nuclei, by comparison of their diagnostic spectra with model
predictions (Kewley et al 2001, Kauffmann et al 2003, Kewley et al 2006).

It is important to note, however, that such diagnostic tools were
developed for spatially unresolved observations where the integrated
emission from the whole galaxy or nebula was contained in a given emission line
ratio. More recently the same tools have been applied for the
interpretation of spatially resolved observations, e.g. IFU
observations of extended HII regions (Garc{\'{\i}}a-Benito et al. 2010, 
Monreal-Ibero et al 2011, Relano
et al 2010). This approach has to be
taken with caution, given that a line of sight projection of a complex
3D structure, which contains fully ionised as well as neutral regions, is not directly
comparable with the results from 
a radiation bound 1D (slab or spherically symmetric) photoionisation
model. Indeed a 2D projection of a 3D cloud will show pixel-to-pixel
variations of the line ratios corresponding to the local gas
conditions along the line of sight.  

This is shown in Figures 4 and 5 where each pixel on the 2D
projections of our pillars is plotted on typical BPT diagrams. The
integrated value for all pixels is shown as the red point on the
figures. The solid red line shown separates the area of the BPT
diagrams that is traditionally considered to be occupied by HII
regions dominated by photoionisation (below the line) from that
occupied by AGNs dominated by shock-ionisation (Kewley et al 2001, Kewley et al
2006). The dashed line in Figure 5 is taken from Kauffmann et al 2003 and represents a stricter partition of the diagram in that it divides pure star--forming regions (below the line) from composite objects such as Seyfert--HII galaxies whose spectra exhibit strong contributio<ns from star formation \textit{and} AGN. The points are
colour-coded by distance from the left-hand side of the
simulation box on which the stellar radiation field impinges. In
particular the green points are for $y$ $>$ 2.6pc, red for 2.6pc $>$ $y$
$>$ 2.2pc, blue for 2.2pc $>$ $y$ $>$ 1.8pc and yellow for $y$ $<$ 1.8pc.
The majority of the points lie, as expected, in the photoionisation
region of the diagram. However, in Figure 4 4$\%$ and in Figure 5 16$\%$ of the points lie well beyond the
red line, apparently suggesting that shocks may be responsible for the observed line
ratios. This is of course not the case in our simulations and the
observed values are simply due to the complex 3D distribution of the
gas, whereby the lines of sight at
these large $y$-positions intersect regions of neutral or quasi-neutral
gas. In the ionised--to--neutral transition regions, singly ionised N
and S are relatively more abundant than in the region where hydrogen is fully ionised
(the ionisation potential for N$^0$ is 14.5eV and that for
S$0$ is 10.3eV), pushing the line ratios to the right of the
diagrams. The message to be taken away from these figures is 
that care should be taken when interpreting spatially resolved
observations using diagnostic diagrams or 1D photoionisation
calculations, which may lead to incorrect conclusions with regard to spatial
variations of abundances and/or the ionisation mechanism (see also e.g. O'Dell
et al 2011 versus Wood et al 2011, or see Balick et al 1994 versus
Goncalves et al 2006). 

\section{Conclusions}

We have presented synthetic observations of optical emission lines
from complex density and velocity fields obtained by the three-dimensional smooth
particle hydrodynamics simulations of Ercolano \& Gritschneder
(2011). These calculations consist of a turbulent box irradiated by a
plane-parallel ionising field, and present at 500kyr density structures reminiscent of nearby
star-forming regions (e.g. the Pillars of Creation, the Horse Head
Nebula). We have performed three-dimensional radiative transfer and
photoionisation calculations and produced emission line maps in
typical ionised gas tracers, including H$\beta$, [OIII]$\lambda$5007, [NII]$\lambda$6584 and
[SII]$\lambda$6717 The resulting composite three-colour images
that we obtained are directly comparable with, e.g., images from the
Hubble Space Telescope (Hester et al. 1996). The morphological appearance of our synthetic pillar images is in good
agreement with the observations. We have also produced spatially
resolved emission line profiles, which show typical FWHM of 2--4 km s$^{-1}$
km s$^{-1}$ and dispersions of 30 km s$^{-1}$, in agreement with photoionised gas with sound-speeds of
approximately 10 km s$^{-1}$. 

Finally we have studied the spatial variation of emission line diagnostics by
means of classical BPT diagrams. The volume-averaged diagnostics are
consistent with the loci expected for photoionised gas, however we
show that significant location-dependent variations of the diagnostic
values are to be expected, due to the complex three-dimensional
ionisation, excitation and temperature distributions along the different lines of
sight. In particular we draw attention to a number of surface elements
that show diagnostics consistent with gas that is shock-ionised,
rather than photoionised, according to the classical interpretation of
the BPT diagrams by means of one-dimensional photoionisation
calculations (e.g. Kewley et al 2006). The gas in this regions
however is not shocked, and the altered diagnostic values are only a
result of lines of sight that intercept large portions of the ionised
to neutral transition regions in the box. We therefore conclude that the
straightforward application of diagnostic diagrams based on
one-dimensional photoionisation calculations is not suitable to the
interpretation of spatially resolved observations of complex
star-forming regions.

\section{Acknowledgements}
We thank the anonymous referee for a constructive report which helped
us to make the paper clearer. M.G. acknowledges funding by the China
National Postdoc Fund Grant No. 20100470108 and the National Science
Foundation of China Grant No. 11003001.

\end{document}